\documentclass[12pt,a4paper]{article}

\usepackage[english]{babel}
\usepackage[dvips]{graphicx,color}

\begin{document}

\title{\bf
Molecular dynamics of C-peptide of ribonuclease A studied by
replica-exchange Monte Carlo method and diffusion theory}

\author{
Giovanni La Penna$^{a,}$\footnote{\ \ e-mail: lapenna@ge.ismac.cnr.it}\ ,
Ayori Mitsutake$^{b,}$\footnote{\ \ e-mail: ayori@rk.phys.keio.ac.jp}\ , 
Masato Masuya$^{c,}$\footnote{\ \ e-mail: masatom@cc.kagoshima-u.ac.jp}\ ,
Yuko Okamoto$^{d,e,}$\footnote{\ \ e-mail: okamotoy@ims.ac.jp}
}
  
\maketitle

\noindent
 $^a$ National Research Council,
 Institute for Macromolecular Studies,
 Section of Genova,
 Via De Marini 6, 16149 Genova, Italy\\
 $^b$ Department of Physics, Faculty of Science and Technology, Keio University,
Yokohama, Kanagawa 223-8522, Japan\\
 $^c$ Computing and Communications Center, Kagoshima University, Kagoshima 890-0065, Japan\\
 $^d$ Department of Theoretical Studies,
 Institute for Molecular Science,
 Okazaki, Aichi 444-8585, Japan\\
 $^e$ Department of Functional Molecular Science,
 The Graduate University for Advanced Studies,
 Okazaki, Aichi 444-8585, Japan\\

\begin{abstract}
\setlength{\baselineskip}{20pt}
Generalized-ensemble algorithm and diffusion
theory have been combined in order to compute 
the dynamical
properties monitored by nuclear magnetic resonance experiments
from efficient and reliable evaluation of statistical averages.
Replica-exchange Monte Carlo simulations have been performed
with a C-peptide analogue of ribonuclease A,
and Smoluchowski diffusion equations have been applied.
A fairly good agreement between the calculated and
measured $^1$H-NOESY NMR cross peaks has been obtained.
The combination of these advanced and continuously improving
statistical tools allows the calculation of a wide variety
of dynamical properties routinely obtained by experiments.
\end{abstract}

%\newpage

%\baselineskip=0.8cm

\begin{section}{Introduction}
In the study of protein folding, a crucial step
is the understanding of secondary structure formations.
The construction of $\alpha$-helices and $\beta$-sheets
from disordered structures and their interconversion is mainly
driven by hydrophobic effects combined with dispersive
interactions and intramolecular hydrogen-bond 
formations~\cite{pala,beta}. Salt bridges and strong electrostatic
interactions can either compete with this driving force
or assist the secondary structure formation, by introducing
long-range average distance constraints.
The latter can be influenced by the characteristics
of the solvent and/or ions eventually enclosed in 
pre-folded configurations.

In order to address these complications related to
the sequence variability in real proteins' secondary
structures, many experiments are performed on relatively
short peptide fragments that {\em{in vitro}} adopt conformations
similar to those revealed in the entire proteins.
These fragments, therefore, may be considered as
independent modules~\cite{moduli} of the original protein
\cite{module2}. 
These experiments allow a great
simplification, replacing most
of the protein matrix with the solvent, while keeping
in the molecule the relevant interactions. On the other hand,
short peptides visit many conformations, and thus 
the usage of statistical tools 
and computer simulations
is required for
the interpretation of experimental data.

Computer simulations in canonical ensemble based on Boltzmann 
weight factor, however, tend to get trapped
in states of energy local minima, and 
it is very difficult to obtain accurate statistical averages
even for small peptide systems.
Generalized-ensemble algorithms are based on artificial,
non-Boltzmann weight factors and perform random walks in
potential energy space, which efficiently alleviates the
multiple-minima problem (for a recent review, see Ref. \cite{GEArev}).
After a single production run, one can calculate accurate
canonical-ensemble averages for a wide range of temperatures
\cite{GEArev}.
Because of the very construction of the generalized-ensemble
algorithms, however, the information of molecular dynamics is lost,
and only static average values can be obtained by these methods.

Nuclear magnetic resonance (NMR) 
has become
an attractive technique because it allows one to monitor
both structural statistics (average distances) and molecular
dynamics (stochastic rotation of vectors) at an atomistic
level and in different solvent/temperature conditions.
These features are, however, intimately linked together in
the experimental data, particularly in those data
routinely measured to obtain structures, {\em{i.e.}},
the $^1$H-NOESY cross peaks (CPs).

In order to calculate dynamical properties from NMR
experiments, a diffusive model has been
designed and casted in a Smoluchowski diffusion equation.
This equation has been solved by matrix expansion methods
and by using a mode-coupling 
approximation~\cite{Fausti2001,LaPenna_piso10}.
This procedure allows the description of
the time-correlation functions (TCFs) that govern
NMR experiments through the computation of a suitable
set of configurational averages.

In this Letter we propose to combine the above two powerful
computational approaches: generalized-ensemble algorithm
and diffusion theory, which allows rigorous calculations
of molecular dynamics inferred by NMR experiments.

A C-peptide analogue of ribonuclease A is here
considered.
The C-peptide of ribonuclease A is one of the smallest peptides that
is known to form $\alpha$-helix conformations and
has been extensively studied by circular dichroism 
(CD) \cite{Buzz} and NMR 
\cite{Osterhout,Bruschw}
spectroscopies.
The peptide has also been studied by computer simulations
\cite{KONF2}--\cite{HO99}.
Nevertheless, a 
quantification of the conformational
population and of its effect on the experimental data was
not possible and the standard methods used to analize NMR
relaxation data cannot be applied~\cite{Lipari}.
We employed one of the commonly used generalized-ensemble algorithms, 
replica-exchange Monte Carlo (REMC) \cite{RE1}, to calculate
statistical averages and various dynamical quantities were 
successfully obtained
by the diffusion theory. 

In section 2 the two mehotds are summarized and computational details are given.
In section 3 the results of the statistics and the NMR CPs
are discussed in a unique frame.
In section 4 conclusions and perspectives are presented.
\end{section}

\begin{section}{Methods}

\begin{subsection}{Replica-exchange method}
We first briefly review the replica-exchange method (REM)
\cite{RE1} (see, for instance,  Refs.~\cite{REMD,GEArev} for details).

The system for REM consists of
$M$ {\it non-interacting} copies (or, replicas)
of the original system in the canonical ensemble
at $M$ different temperatures $T_m$ ($m=1, \cdots, M$).
We arrange the replicas so that there is always
exactly one replica at each temperature.
Then there is a one-to-one correspondence between replicas
and temperatures.
Let $X = \left\{\cdots, x_m^{[i]}, \cdots \right\}$
stand for a ``state'' in this generalized ensemble.
Here, $x_m^{[i]}$ stands for the state of the $i$-th replica
(at temperature $T_m$); 
the superscript $i$ and the subscript $m$ in $x_m^{[i]}$
label the replica and the temperature, respectively.
Each state $x_m^{[i]}$ is specified by the coordinates $q^{[i]}$ (and momenta $p^{[i]}$)
of all the atoms in replica $i$.

A simulation of REM
is then realized by alternately performing the following two
steps.
Step 1: Each replica in canonical ensemble of the fixed temperature
is simulated $simultaneously$ and $independently$
for a certain MC or MD steps.
Step 2: A pair of replicas,
say $i$ and $j$, which are
at neighboring temperatures $T_m$ and $T_n$, respectively,
are exchanged:
$X = \left\{\cdots, x_m^{[i]}, \cdots, x_n^{[j]}, \cdots \right\}
\longrightarrow \
X^{\prime} = \left\{\cdots, x_m^{[j]}, \cdots, x_n^{[i]},
\cdots \right\}$.
The transition probability of this replica exchange is given by
the Metropolis criterion:
\begin{equation}
w(X \rightarrow X^{\prime}) \equiv
w\left( x_m^{[i]} ~\left|~ x_n^{[j]} \right. \right)
= {\rm min} \left( 1, \exp \left( - \Delta \right) \right)~, 
\label{eqn24}
\end{equation}
where
\begin{equation}
\Delta = \left(\beta_m - \beta_n \right)
\left(E\left(q^{[j]}\right) - E\left(q^{[i]}\right)\right)~.
\label{eqn23b}
\end{equation}
Here,
$E\left(q^{[i]}\right)$ and $E\left(q^{[j]}\right)$
are the potential energy of the $i$-th replica
and the $j$-th replica, respectively.
In the present work we employ Monte Carlo algorithm
for Step 1.
When the potential energy depends on temperature as in the
present case (see Eq. (\ref{eqtote}) below),
we should use the following $\Delta$ instead of that in Eq. (\ref{eqn23b})
\cite{MKOH} (see also \cite{MREM}):
\begin{equation}
\Delta = \beta_m
\left(E\left(q^{[j]};T_m\right) -
E\left(q^{[i]};T_m\right)\right)
- \beta_n
\left(E\left(q^{[j]};T_n\right) -
E\left(q^{[i]};T_n\right)\right)~.
\label{Eqn21}
\end{equation}

A random walk in ``temperature space'' is
realized for each replica, which in turn induces a random
walk in potential energy space.  This alleviates the problem
of getting trapped in states of energy local minima.

The canonical expectation value of a physical quantity $A$
at temperature $T_m$ ($m=1, \cdots, M$) can be calculated
by the usual arithmetic mean as follows:
\begin{equation}
<A>_{T_m} = \frac{1}{N_{sim}} \sum_{t=1}^{N_{sim}}
A\left(x_{m}^{[i]}(t)\right)~,
\label{eqn17}
\end{equation}
where $N_{sim}$ is the total number of measurements made
at temperature $T_m$.
Note that the above summation is taken over different replicas $i$
($i=1, \cdots, M$) 
that happens to correspond to the fixed temperature $T_m$ 
at the moment of measurement.
The expectation values at any intermediate temperature can, 
in principle,
be calculated by the multiple-histogram reweighting 
techniques \cite{REMD}, but in this article we limit our discussions to
the above $M$ temperature values.
\end{subsection}

\begin{subsection}{Calculation of NMR parameters}
In this subsection we briefly review recent advances of diffusion
theory~\cite{Fausti2001,LaPenna_piso10}
that are applied to the calculation of the $^1$H-NOESY NMR cross peak
intensities (CPs).

The CP intensities $O_{h,k}$ at mixing time $t_m$ 
between spin $h$ and spin $k$
can be calculated by the following equation:
\begin{equation}
O_{h,k} \left( t_m \right) = \frac{R_{h,k}}{\Delta} \ \exp 
 \left( { - \sigma \ t_m} \right) \ \sinh \left( {\Delta \ t_m} \right)~,
\label{eq_singleCP}
\end{equation}
where $R_{h,k}$, $\Delta$, and $\sigma$ are all functions of the
following spectral densities
(see Refs.~\cite{Fausti2001,LaPenna_piso10} for details):
\begin{equation}
J_{h,k}(\omega ) = 
  2\int\limits_0^\infty  {\cos \left( {\omega t} \right)} \;TCF_{h,k}\left(
 t \right)dt~.
\label{eq_specdens}
\end{equation}
Here, $TCF_{h,k}$ is a time-correlation function
of 2nd-rank tensor components of the vectors ${\bf{r}}_{h,k}$ 
joining proton $h$ and proton $k$.
These TCFs at temperature $T$ have the form~\cite{Cavanagh}
\begin{equation}
TCF(t) = \sum\limits_{M =  - 2}^2
 {\left\langle {\left[ {D_{M,0}^{(2)*} (\Omega (t))} ~/r(t)^3 \right]\left[ {D_{M,0}^{(2)}
 (\Omega (0))} ~/r(0)^3 \right]} \right\rangle}_T~,
\label{eq_TCF}
\end{equation}
where $D_{M,0}^{(2)}$ are irreducible spherical
tensors~\cite{Rose}, and $\Omega$ and
$r$ are the direction and the modulus, respectively,
of the given H-H vector involved in the NOESY CP (subscripts $h$ and $k$
are henceforth suppressed for clarity).

In order to separate the effect
of the modulus from the orientation of the unit vector (direction)
in the above TCF, we also calculate the orientational TCF from
\begin{equation}
TCF_O(t) = \sum\limits_{M =  - 2}^2
 {\left\langle {\left[ {D_{M,0}^{(2)*} (\Omega (t))} \right]\left[ {D_{M,0}^{(2)}
 (\Omega (0))} \right]} \right\rangle}_T  = P_2(\cos[\theta(t)])~,
\label{eq_TCFo}
\end{equation}
where $P_2$ is the 2nd-order Legendre polynomial.
The orientational mobility can be described by the correlation time $\tau$
that is the integral of TCF$_O$:
\begin{equation}
\tau = \int\limits_0^\infty  { TCF_O\left( t \right)~dt}~.
\label{eq_tauc}
\end{equation}

The mode-coupling diffusion (MCD) theory of the dynamics of
a biological macromolecule
in solution is adopted for the computation of the above TCFs
of Eqs.~(\ref{eq_TCF}) and (\ref{eq_TCFo}).
The MCD approach~\cite{LaPenna_rig,Perico_Pratolongo} can
be briefly summarized as follows. Given a polymer of $N_a$
beads of friction coefficients $\zeta_i$ and coordinates ${\bf{r}}_i$,
connected by $N_b$ bonds $({\bf{l}}_i, i=1,...,N_b)$,
the dynamics of each variable ${\bf{l}}_i$,
is regulated by the
operator $L$, adjoint to the diffusion Smoluchowski operator $D$:
\begin{equation}
\frac{{\partial {\bf l}}}{{\partial t}} = L{\bf l};
 \quad L = \sum\limits_{i,j = 1}^{N_a } {\left[ {\nabla _i {\bf D}_{i,j} \nabla _j
           - \left( {\nabla _i U/k_B T} \right){\bf D}_{i,j} \nabla _j } \right]}~,
\label{eq_Langevin}
\end{equation}
where $U$ is the potential energy of the beads as a
function of the bead coordinates, $k_B$ is the Boltzmann constant, $T$
is the absolute temperature, and {\bf{l}} is the $3 \times N_b$ dimensional
array containing all the bond vectors ${\bf{l}}_i$.

By expanding the conditional probability (solution to the Smoluchowski
equation) in a complete set
of eigenfunctions of $L$, the time autocorrelation function (TCF) of
any coordinate-dependent
dynamic variable with zero average $f(t)$ may be expressed in
the standard form
\begin{equation}
\langle {\kern 1pt} f(t)\;f(0)\rangle_T \;\; = \;\;\sum\limits_i \, \langle {\kern
 1pt} f\,\,\psi _i \rangle_T \;\langle \psi _i \,\,f{\kern 1pt} \rangle_T \;{\rm e
xp}\,{\rm (} - \lambda _i \,t{\rm )}~,
\label{eq_ff}
\end{equation}
where $-\lambda_i$ and $\psi_i$ are respectively the eigenvalues and the
normalized eigenfunctions of the operator $L$:
\begin{equation}
L\psi _i  =  - \lambda _i \psi _i ~.
\label{eq_evec}
\end{equation}

This eigenvalue equation becomes a matrix equation, with
the matrix elements being equilibrium averages at temperature $T$
(see Refs.~\cite{Fausti2001,LaPenna_piso10} for details).
In the present work, we use the replica-exchange Monte Carlo
method for the calculation of these averages (see Eq. (\ref{eqn17})).

\end{subsection}

\begin{subsection}{Computational details}
 The configurational statistics of a C-peptide analogue
\cite{Osterhout}
with the amino-acid sequence AETAAAKFLRAHA and uncharged N- and C- termini
have been simulated by replica-exchange Monte Carlo method.
Residue His 12 was protonated in order to better match the
NMR experimental conditions of pH 5.2.  Other charged residues were
Glu 2$^-$, Lys 7$^+$, and Arg 10$^+$. The number of atoms in the model
was 195.
    
The total ``potential energy'' function 
$E(q;T)$
that we used is the sum of
the conformational energy term of the solute $E_P(q)$ and
the solvation free energy term $E_{SOL}(q;T)$
for the interaction of the peptide with
the surrounding solvent:
\begin{equation}
E(q;T) =E_{P}(q) + E_{SOL}(q;T)~.
\label{eqtote}
\end{equation}
The parameters in the conformational energy as well as the molecular geometry 
were taken form ECEPP/2 \cite{ECEPP2}.
The sigmoidal, distance-dependent
dielectric function of Ref.~\cite{O2} was used.  
    
The solvation free energy that we used is given by a linear combination
of the solvent-accessible surface area (SASA) $A_i$
of each non-hydrogen atom $i$:
\begin{equation}
E_{SOL}(T_0) = \sum_{i} \sigma_i A_i~,
\label{eqasa}
\end{equation}
where $\sigma_i$ are the proportionality constants, $T_0 = 298$ K,
and the dependence on the coordinates $q$ is now suppressed.  
The temperature dependence of the solvation free energy
was taken into account, following the prescription in Ref.~\cite{Ooi}:
\begin{equation}
E_{SOL}(T) = \frac{T}{T_0} E_{SOL}(T_0)
+ H_{SOL}(T_0) \left(1 - \frac{T}{T_0}\right)
- C_{SOL}(T_0)\left[T\ln\left(\frac{T}{T_0}\right) + T_0 - T\right]~,
\label{eqasaT}
\end{equation}
where
$H_{SOL}$ and $C_{SOL}$ are enthalpy and heat capacity, respectively
\cite{Ooi}.
  
The SASA was calculated by the computer
code NSOL \cite{NSOL}.
The computer code KONF90 \cite{KONF2} was used, 
and MC simulations based on REM were performed.
In each Monte Carlo sweep all the independent dihedral angles
except for the peptide-bond dihedral angles $\omega$, which were 
fixed at 180 degrees, were updated once and
the Metropolis test was performed for each update. The number of changeable
torsion angles was then 55 and the maximum torsional
change was $\pm 180$ degrees. 

For REMC we used 10 replicas.
The corresponding temperatures were 200, 233, 276,
317, 370, 432, 504, 588, 686, and 800 K.
These temperatures were chosen to span the temperature
range between 200 and 800 K and to contain the temperature
of the NMR experiments (276.15 K). 
The initial conformations were randomly generated.
The replica exchange was tried every 20 MC sweeps. 
For an optimal performance of REMC simulations, the acceptance
ratios of replica exchange should be sufficiently uniform and large
(say, $>$ 10 $\%$ ).  The acceptance ratio was indeed found to be 
in the range 14--24 \%, and we observed that
each replica underwent an unbiased random walk in the potential energy space
(and that each temperature underwent a random walk in the replica
space). 
After 1,100,000 MC sweeps of
equilibration, the REMC production run of 3,000,000 MC sweeps for
each replica was made.
The configurations were stored every 10 MC sweeps for data analyses.
This amounts to 300,000 configurations for each temperature
(or each replica).

As for the diffusion equation, the first step is
to approximate atoms or group of atoms in the molecule
as friction points. In the present model, friction
points were located on 55 heavy atoms (beads) among the total of 195
atoms. 
The friction was computed by using Stokes' law
with stick boundary conditions with Stokes' radii
obtained summing the accessible surface area to a spherical probe
of 0 radius (ASA0) of the atoms grouped in the bead~\cite{Pastor_Karplus}.
The Stokes' radii ranged from 0.09 nm (C$_\alpha$ in all residues
that include H$_\alpha$ only) to 0.24 nm (the last portion
of Arg 10 side chain, that includes N$_\epsilon$, C$_\zeta$, 
N$_{\eta1}$, N$_{\eta2}$ and
all the bonded hydrogen atoms).
The water viscosity $\eta$ was 0.001 Pa s and it 
was assumed independent of temperature.

The second step is to assess the convergence of
the used basis set in solving the eigenvalue equation
for the adjoint of the diffusion operator.
 Both the long-time sorting procedure (LTSP)~\cite{Freed} and the 
maximum correlation approximation (RM2-II basis set of MCA)~\cite{LaPenna_piso10}
were applied to select the most important terms of the 
infinite mode-coupling basis set.
The results coupling the five 1st-order
lowest-rate modes for 1st-rank variables
up to the 2nd-order for 2nd-rank variables (MCA with $e=5$
in the notations of Ref.~\cite{LaPenna_piso10})
were almost identical with the 2nd-order LTSP using up to
600 basis functions (data not shown). Therefore,
the MCA basis set built with $e=5$ (240 basis functions) 
was used for all of the following calculations.

No significant changes
in the ten lowest-rate 1st-order/1st-rank relaxation modes
were observed reducing the number of configurations
from 300,000 down to 10,000.
Therefore, for the calculations of the statistical averages required
to solve the diffusion equation and to compute
the TCFs at each temperature, 10,000 of the 300,000 
recorded configurations were used. 
\end{subsection}

\end{section}

\begin{figure}
\includegraphics{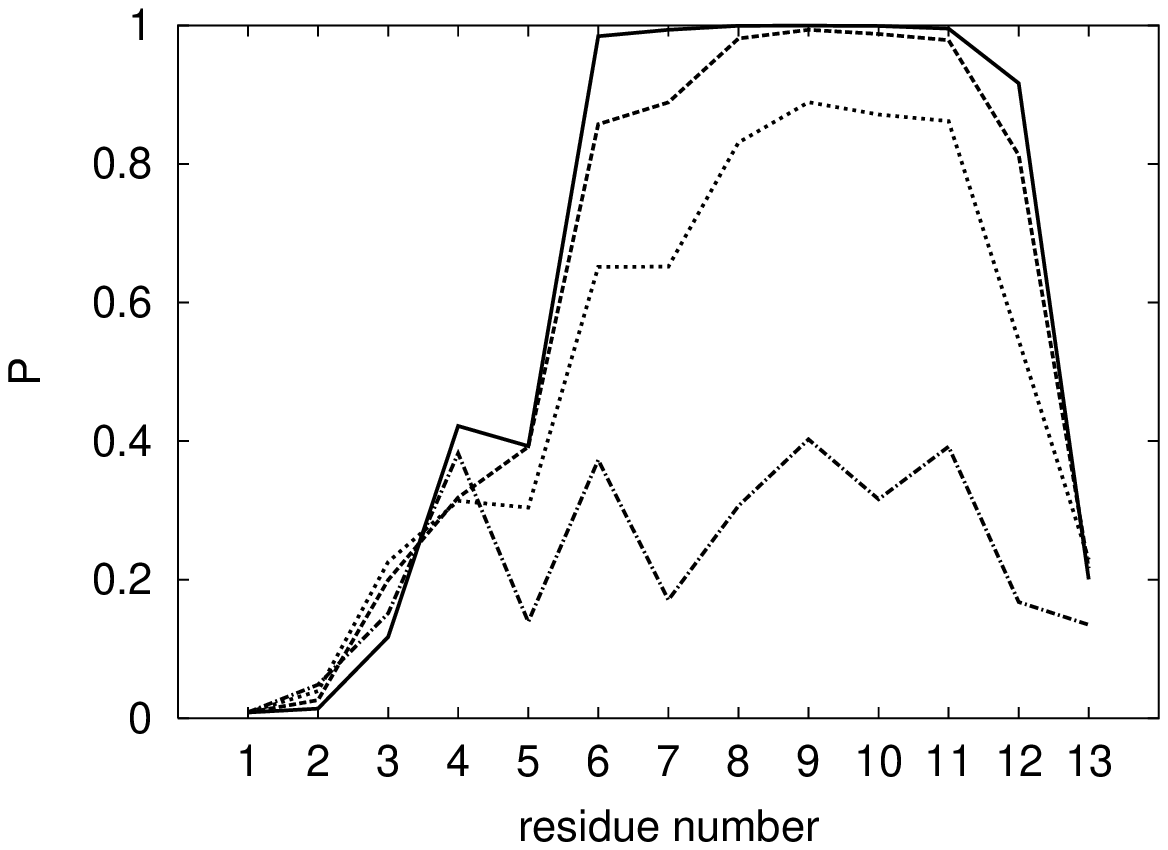}
\vspace{2cm}
\caption{La Penna et al.}
\label{fig_helr}
\end{figure}

\begin{figure}
\includegraphics{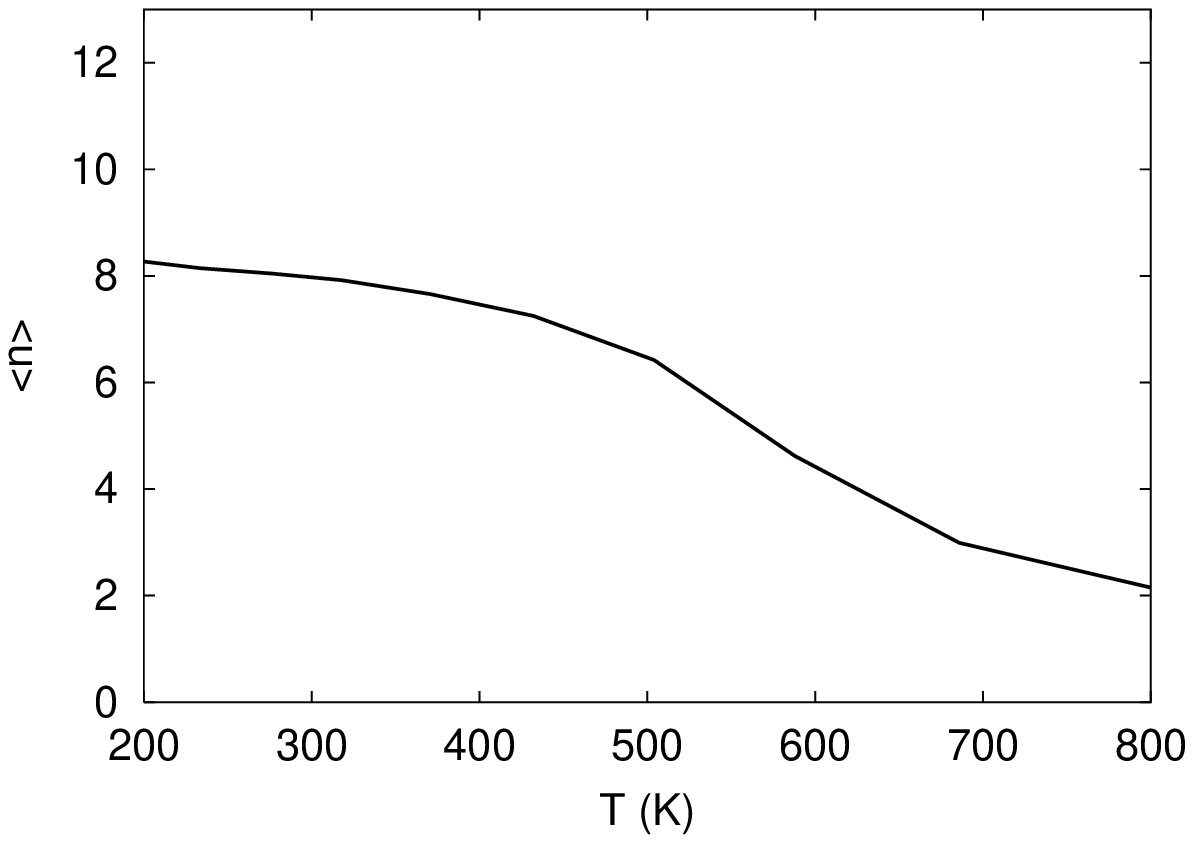}
\vspace{2cm}
\caption{La Penna et al.}
\label{fig_hel}
\end{figure}

\begin{figure}
\scalebox{0.6}
{
\includegraphics{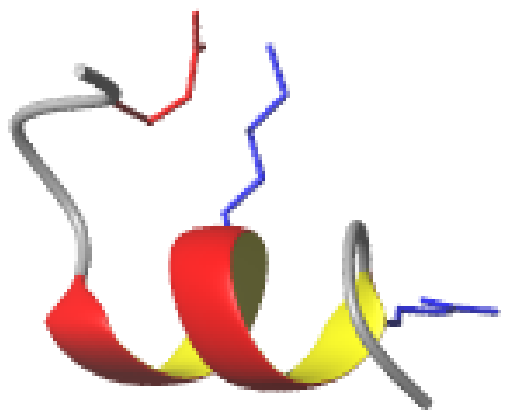}
\hspace{0.5cm}
\includegraphics{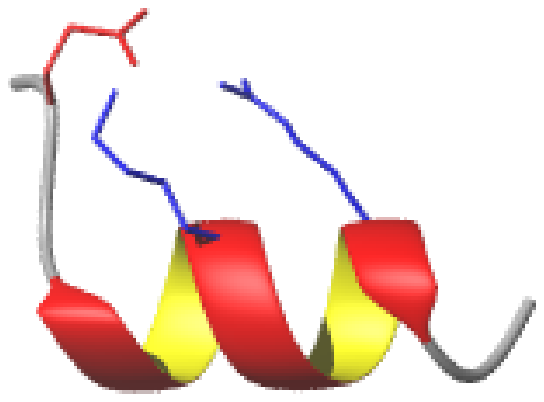}
\hspace{0.5cm}
\includegraphics{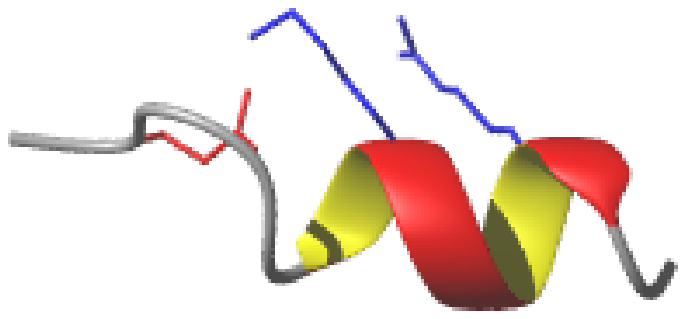}
}

\vspace{1.0cm}
\includegraphics{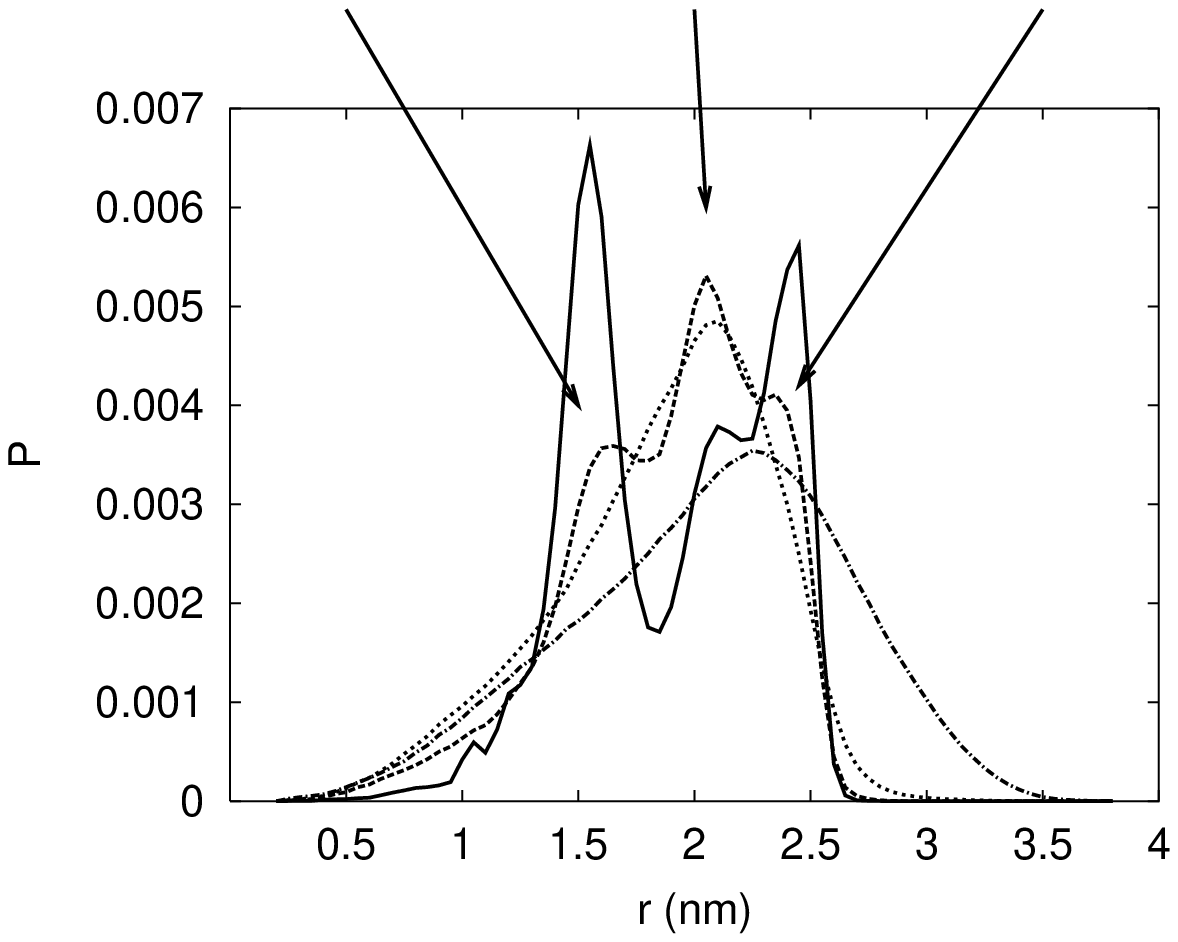}
\vspace{2cm}
\caption{La Penna et al.}
\label{fig_ee}
\end{figure}

\begin{section}{Results and Discussion}
The C-peptide is relatively rich in hydrophobic residues
(Ala, Leu, and Phe) and, therefore, in a water environment
is expected to be mainly in $\alpha$-helix conformation.
On the other hand, the presence of charged residues, namely,
Glu 2$^-$ (near the N-terminus), 
Lys 7$^+$, Arg 10$^+$, and His 12$^+$ (near the C-terminus),
will have significant effects on the conformational
states of the peptide.
In Fig.~\ref{fig_helr} the average $\alpha$-helicity
as a function of residue number (or, probability of each
residue being in the $\alpha$-helix state) is shown for four
different temperatures (276 K, 370 K, 504 K, and 686 K).
Here, we considered
that a residue is in the $\alpha$-helix state
when the backbone dihedral angles ($\phi,\psi$) fall in
the range ($-70 \pm 30^{\circ},-37 \pm 30^{\circ}$), and
Eq. (\ref{eqn17}) was used to calculate the average helicity.
At the lowest temperature ($T=276$ K) among the four,
residues 4-12 are in helical state (especially, residues
6-11 are completely helical), which is exactly the same
location of $\alpha$-helix as found in the corresponding
structure from the X-ray experiments of the entire
ribonuclease A \cite{Xray}.  Helicity decreases as the
temperature is raised because of the increased thermal fluctuations.
High helicity persists up to $T=504$ K (especially in
residues 6-11), and finally at the highest temperature among the four 
($T=686$ K), extended helical conformation ceases to exist.
Note also that the N-terminus is not helical even at
276 K in agreement with the NMR experiments
\cite{Osterhout}.
  
Fig.~\ref{fig_hel} shows the behaviour of the average
total helicity (or, average total number of helical
residues) as a function of temperature. 
As was observed in Fig. 1, the residues near the N-terminus
are rarely in helical state, and this is the reason why even
at the lowest temperature ($T=200$ K) the average total number
of helical residues is only about 8.
The slight decrease of helicity between 370 K and 500 K
is due to residues 6, 7, and 12 that lose $\alpha$-helical
population, while the further decrease beyond 500 K
involves the demolition of the remaining 
$\alpha$-helix in residues 8-11. The disorder in configurational statistics beyond 600 K
still keeps part of the electrostatic interactions
characterizing the low-energy structures (see discussion below),
thus representing a stiff disordered polymer segment.
Note that our statistics show that high total helicity ($> 50$ \%)
persists as high as $T \approx 500$ K, while the experiments
observe high helicity only near $T=273$ K \cite{Buzz,Osterhout}.
This shift in helix-coil transition temperature is presumably
due to the fact that our energy functions including the solvent model
are not accurate enough to reproduce the absolute temperature
dependence of experiments.  As discussed below in detail, our
simulation results around $T=400$ K best reproduce the NMR
experiments (which were conducted at $T=276$ K).

In Fig.~\ref{fig_ee} the end-to-end distance
distribution is shown at the same four temperatures as in Fig.
\ref{fig_helr}.
Here, the end-to-end distance 
is defined to be the distance between
N of Ala-1 and O of Ala-13.
At the two lower temperatures ($T=276$ K and 370 K)
we observe three peaks in the
distributions, which suggests that there exist three groups
of similar conformations.
We refer to the three groups as Groups 1, 2, and 3 from
left to right in the Figure.
Representative conformations from
each group (the lowest-energy conformation in each group)
are also shown in
Fig.~\ref{fig_ee}.
All these three groups of conformations have a common
$\alpha$-helix structure in residues 5-11.
The end-to-end distance is about 1.5 nm, 2 nm, and 2.5 nm
for Group 1, Group 2, and Group 3, respectively.
Conformations in Group 1 are characterized by a salt bridge
between side chains of Glu 2$^-$ and Lys 7$^+$ and a
bend towards the N-terminus so that the end-to-end distance
is the shortest among the three groups.  Conformations in
Group 2 are charecterized by two salt bridges between
Glu 2$^-$ and Lys 7$^+$ and between Glu 2$^-$ and Arg 10$^+$.
Note that this group has the most similarity to the X-ray
structure \cite{Xray}, which also has the 
Glu 2-Arg 10 salt bridge.  The backbone root-mean-square
distance of the Group 2 structure in Fig. 3 from that
of X-ray experiments is 0.12 nm, while those of Group 1
and Group 3 are 0.32 nm and 0.19 nm, respectively.  Finally,
conformations of Group 3 also have a salt bridge between
Glu 2$^-$ and Lys 7$^+$.  However, the N-terminus is pushed 
away from the $\alpha$-helix and the structure is rather
extended.

As is shown in Fig. 3, the highest populated peak at the lowest
temperature (276 K) corresponds to Group 1.
As temperature increases (370 K), this peak decreases
in population and the peak of Group 2
increases,
thus showing an increased stability of a longer $\alpha$-helical
segment (characterized by a larger end-to-end distance).
At the third temperature (504 K) only a single peak
(of Group 2) exists, which suggests that the ``native-like''
structure (i.e., Group 2) is the most stable among the
three groups.
At the highest temperature (686 K) 
we have a single peak at a different end-to-end distance (about 2.2 nm),
which corresponds to a coil structure.

The above results can be explained in terms
of microscopic interactions.
Glu 2 is involved in salt bridges with Lys 7 and, less frequently, with Arg 10.
These salt bridges bend the N-terminus toward the short
$\alpha$-helical region in residues 6-12 and tend to make the whole
molecule more compact. These interactions are more efficient
at the low temperatures (276 K).
Increasing the temperature, the observed salt bridges become
less populated and the bending is more frequenly released, thus
allowing an increase of the molecular extension and, eventually,
a more extended $\alpha$-helical region including the 
Ala-Ala-Ala sequence.
These extended $\alpha$-helical configurations are actually
present in the statistics and are characterized by an
end-to-end distance of about 2.5 nm.
Beyond 500 K, the hydrophobic interactions mainly
responsible for the $\alpha$-helical stability are
no more effective. On the other hand, the salt bridge
between Glu 2 and Lys 7 is still contained in the
statistics (although it is weaker).

We now study the effects of this conformational distribution 
and of its temperature dependence on the NOESY CPs.
The NOESY experiment was performed
at the temperature of 276.15 K, pH 5.2, in the static magnetic field
corresponding to $\nu (^1H) = 500$ MHz, and using a mixing time of
400 ms. These experimental CPs are reported in reference~\cite{Osterhout}.
 Unfortunately, the experimental results are reported in arbitrary
units and do not allow a quantitative estimate
of global dynamical effects, but it is possible
to analyze the behaviour of CPs for different
proton pairs in the molecule.

The experimental CPs are reported for
three types of proton pairs: HN(i)-HN(i+1) (referred to as NN CPs, hereafter),
H$_\alpha$(i)-HN(i+1) (AN CPs, hereafter), 
and some long-range CPs involving side chains (LR CPs, hereafter).
First, the NN CPs were computed at four
different temperatures that were used in the REMC simulation (i.e., 276,
370, 504 and 686 K). The above analysis of
several statistical quantities, such as the helicity
for each residue at each temperature (Fig.~\ref{fig_helr}) showed
that the molecular model undergoes a helix-coil
structural transition with a transition temperature
estimated between 450 K and 600 K (Fig.~\ref{fig_hel}). 
The real sample undergoes
the same transition at lower temperatures;
the experimental NOESY CPs are not observable even at room temperature.
In Figs.~\ref{fig_CPT}(a) and \ref{fig_CPT}(d) the behaviour of the NN and AN CPs
with temperature is shown.
In both NN and AN data sets, negative CPs are present for
the lowest temperature. The experimental NOESY pulse-sequence
is phase-sensitive and CPs of opposite sign with respect to the
diagonal peaks are not observed at the experimental conditions.
Explanation of the change in sign of the CPs is not trivial.
The explanation in terms of a global $\tau$, which is the inverse
of the unique relaxation rate usually assumed to govern
the 2nd-rank rotation of the given H-H distance vector, is here
not meaningful, because of the high flexibility that allows
many rates to play a significant role in each spectral density.
The change in sign of CPs can be related to the change in
the whole 2nd-rank rate spectrum that is obtained by the 
calculations (data not shown);
at the lower temperature the spectrum is characterized by several
gaps that are progressively smoothed by increasing temperature.
It is expected that for the highest temperature the spectrum
be almost a smooth function of the relaxation mode as in
a polymer random coil. The intermediate situations, where the
internal kinetics is faster, but the molecule is still characterized
by groups of internal modes separated by rate gaps,
can produce the change of sign in CPs and CPs relatively high
in magnitude, as in the experiments. On the other hand,
the large CPs calculated at high temperatures (e.g., 
the results at $T=686$ K in
Fig.~\ref{fig_CPT}) can be related to the limitations
of the basis set construction in the MCA approximation
({\em{i.e.}}, $e$=5).

Even if the orientational mobility of the H-H unit vector and
the H-H average distances are coupled in the TCF of Eq.~(\ref{eq_TCF}),
the behaviour of CPs with temperature can be partially understood in terms 
of the correlation time (Eq.~(\ref{eq_tauc})) and of the average H-H distances.
In Figs.~\ref{fig_CPT}(b) and \ref{fig_CPT}(e) the correlation times for the
NN and AN unit vectors are respectively plotted, and in 
Figs.~\ref{fig_CPT}(c) and \ref{fig_CPT}(f) the average moduli of the same
proton pairs are plotted. Correlation times globally decrease
as is expected. The decrease in orientational rigidity of the 
low-temperature helical region occurs up to the highest temperature where the
difference in orientational rigidity between residues 1-3 and
4-11 is not significant, as it is expected for a non-structured
molecule. On the other hand, it must be noticed that the distances
behave differently: The differences between distances in the two
regions disappear at the highest temperature, but the distances are
still small enough to give contribution to the CPs, especially
the AN CPs. Therefore, the structural information contained
in CPs must be searched in the behaviour of both sets of CPs,
and a separate analysis of AN and NN CPs may be misleading.

The orientational mobility is more sensitive to temperature
than the average distances, the latter being
more stable. However, it is evident that the
approximation of assuming the same orientational
behaviour for all the H-H vectors along the sequence does
not hold;
at 370 K an increase of $\tau$ from 50 ps to 300 ps
moving the H-H vector from the N-terminus deep into
the $\alpha$-helix (Phe 8) can be observed.

As a consequence of the analysis of the temperature
behaviour of CPs, the statistics between 370 K and 504 K
can be considered to reproduce the experimental
conditions at $T=276.15$ K.
A shift in the order-disorder transition temperature
with respect to experiments is expected
and always occurs in other simulated systems like liquid crystals.

In Fig.~\ref{fig_CPexp} CPs at four selected temperatures
are compared with the available
experimental data. All calculated and measured CPs are scaled
by a unique factor in order to have the AN CP of Ala 3, which is the largest
experimental CP available, equal to 1.
The increase of NN CPs from the N-terminus to
the helical region (residues 6-12) is
reproduced by calculations at $T=370$ K, with the largest
deviations beyond Phe 8 (Fig.~\ref{fig_CPexp}(a)). The significant
decrease of AN CPs beyond residue 3 is qualitatively reproduced 
(Fig.~\ref{fig_CPexp}(b)); CPs beyond residue 5 
are slightly smaller than those found in
the experiments and the largest deviations from experiments is found
for Ala 4. This latter deviation is also found in the $^3J$(HN-H$_\alpha$)
coupling constants (data not shown). 
The Coupling
constants have been computed using the Karplus equation
$^3J = 1.9-1.4~\cos(\phi')+6.4~\cos(\phi')$ where
$\phi'$ is the HN-N-C$_\alpha$-H$_\alpha$ dihedral angle.
Residue Ala 4 presents the largest deviations
from experiments, thus implying that the torsional
state in the region 4-5 is not well captured by the
simulation.

It must be noticed that NN CPs are better reproduced
at 370 K, while AN CPs seem to be better reproduced
at 432 K. This suggests that a qualitative reproduction
of both sets of CPs could be achieved at an intermediate
temperature.

The computed LR CPs are all weak; the strongest one
(H$_{\beta}$(Thr 3)-HN(Ala 5) at $T=370$ K) 
is $1.2 \times 10^{-3}$ compared to
the experiment where it is found to be about the same order of magnitude of the
NN CPs. These low CPs are caused by the large average distances, because
the orientational correlation times of LR H-H unit vectors are
in the range 100-200 ps, therefore only slightly smaller than
those of NN H-H unit vectors in the helix (about 200 ps, see 
Fig.~\ref{fig_CPT}(e)). On the other hand, the H$_{\beta}$(3)-HN(5)
average distance is $0.5\pm0.1$ nm, which is too large to produce
even a weak CP.
However, the Glu 2-Lys 7 salt bridge that was suggested by
these long range NMR constraints is found very stable up to
$T=432$ K in the model (0.4 nm), while it becomes 0.5 nm at
the highest analyzed temperature. Therefore, even if the NMR structural details 
are not reproduced in terms of H-H distances, the global features
that are responsible for the most populated conformations are
contained in the model up to the temperature where the comparison
between experimental and computed NN and AN CPs is qualitatively good.

\end{section}

\begin{section}{Conclusions}
 In this work two recent advances in statistical mechanics
have been combined together in the study of the statistics and
dynamics of a small peptide, the C-peptide of ribonuclease A. 
The replica-exchange Monte Carlo simulation has been used
to sample molecular configurations in the canonical ensemble
at several temperatures in the range 200-800 K. This
method, together with other generalized-ensemble algorithms, 
has the advantage of allowing the overtaking of energy barriers,
connecting low temperature and high temperature trajectories.
Diffusion theory in the form of the Smoluchowski equation for
the conditional probability governing the stochastic time evolution
of intramolecular segments' orientation, has then been used
to model the orientational correlation functions and to compute
the $^1$H-NOESY NMR cross peaks that are experimentally available.

The combination of generalized-ensemble statistics and
diffusion theory, frequently updated by technical progresses
making both methods more robust and efficient, allows
the direct calculation of NMR data and other dynamical properties, 
thus closing the gap between
theoretical or computational models and experiments.

\noindent
{\bf Acknowledgements}: \\
This work has been done within the bilateral agreement
for scientific and technological cooperation between
the National Research Council (Italy) and the Japan
Society for the Promotion of Science (JSPS).
The authors thank Angelo Perico (CNR) for many
suggestions on this work.
The simulations were
performed on the computers at the Research Center for
Computational Science, Okazaki National Research Institutes.
This work was supported, in part, by a grant from
the Research for the Future Program of the Japan Society for the
Promotion of Science (JSPS-RFTF98P01101).

\end{section}

\vspace{2cm}
%\newpage

%\noindent
\centerline{\bf Figure Captions}

\begin{itemize}
\item Figure~\ref{fig_helr}.  Average helicity
as a function of residue number at four temperatures: 
$T=276$ K (solid line), $T=370$ K
(dashed line), $T=504$ K (dotted line) and $T=686$ K (dotted-dashed line).

\item Figure~\ref{fig_hel}.  Average total helicity
as a function of temperature.

\item Figure~\ref{fig_ee}.  Distribution of the end-to-end distance at
four temperatures: $T=276$ K (solid line), $T=370$ K
(dashed line), $T=504$ K (dotted line) and $T=686$ K 
(dotted-dashed line); 
the arrows identify the peaks in terms
of representative conformations.  Besides backbone, the side chains of Glu 2, Lys 7, and
Arg 10 are also shown.  The conformations were
drawn with MolMol~\cite{MolMol}.

\item Figure~\ref{fig_CPT}.  NN (a-c) and AN (d-f) CPs, 
correlation times $\tau$,
and average H-H distances as functions of
residue number: $T=276$ K (squares), $T=370$ K
(circles), $T=504$ K (triangles) and $T=686$ K (diamonds).

\item Figure~\ref{fig_CPexp}.  NN (a) and AN (b) CPs as functions of
residue number: experimental data (filled squares) and calculated 
data at $T=276$ K (squares), 
$T=370$ K (circles), 
$T=432$ K (triangles), and $T=504$ K (diamonds).

\end{itemize}

\pagestyle{empty}

\newpage

\newpage

\newpage

\begin{figure}[h!]
\scalebox{0.7}
{
\includegraphics{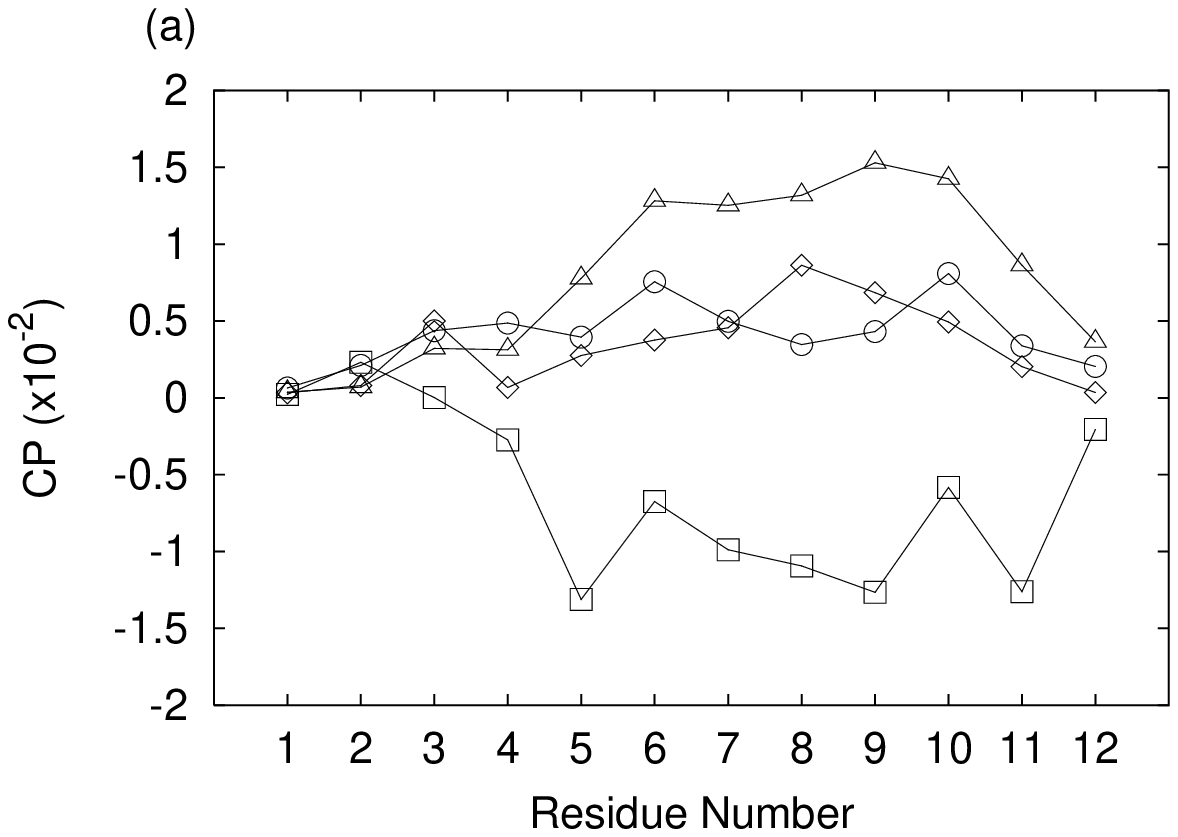}
\hspace{0.2cm}
\includegraphics{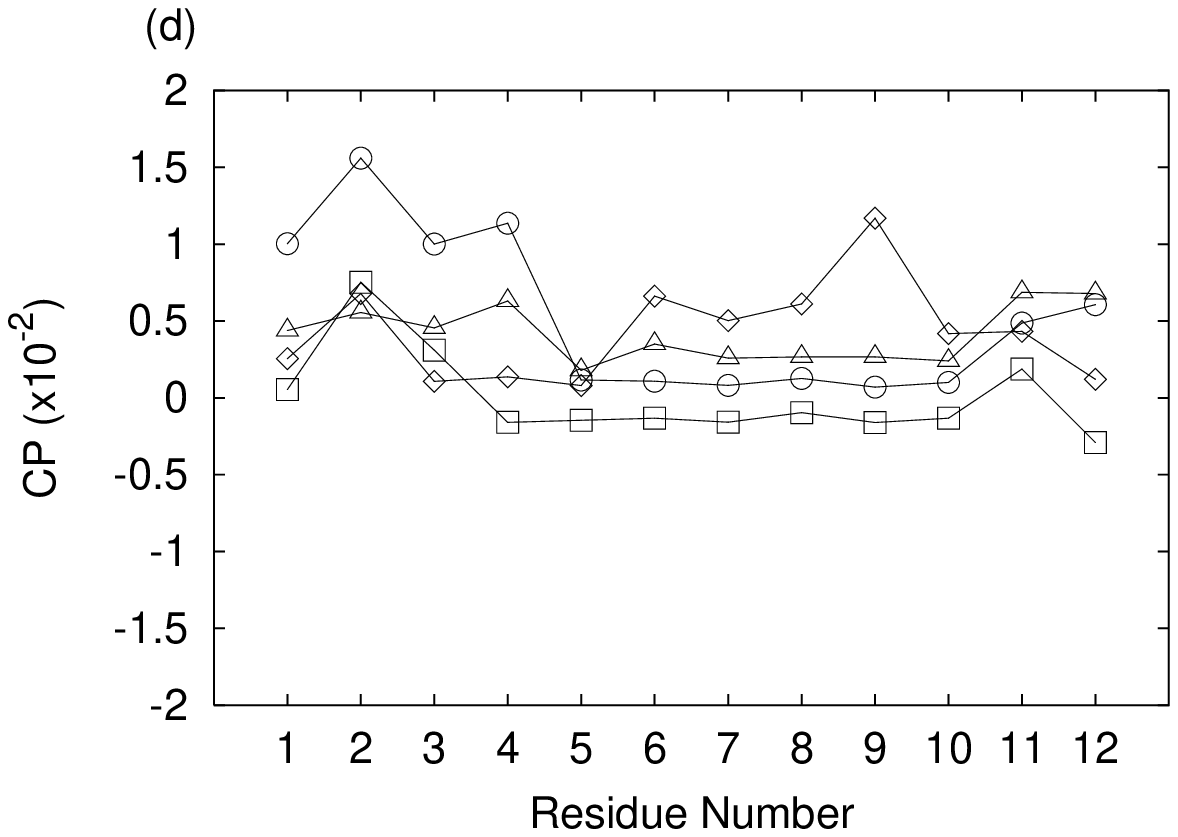}
}

\scalebox{0.7}
{
\includegraphics{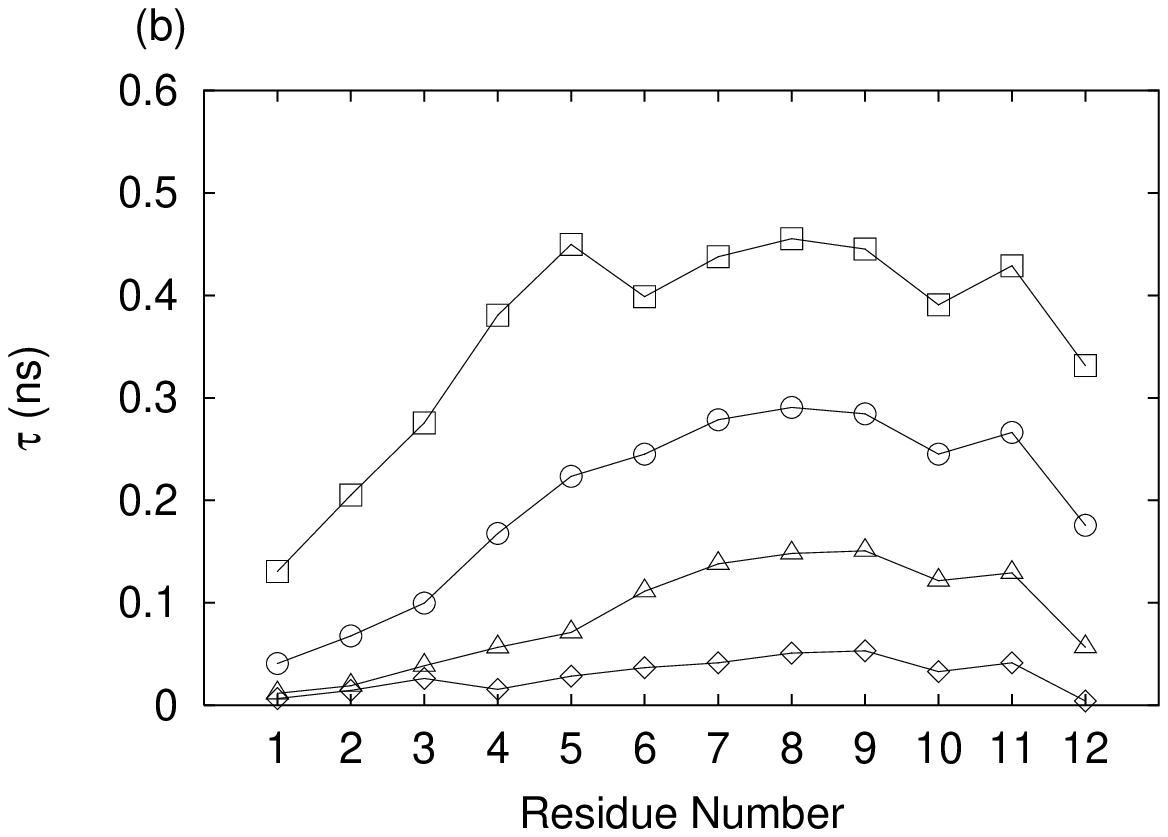}
\hspace{0.2cm}
\includegraphics{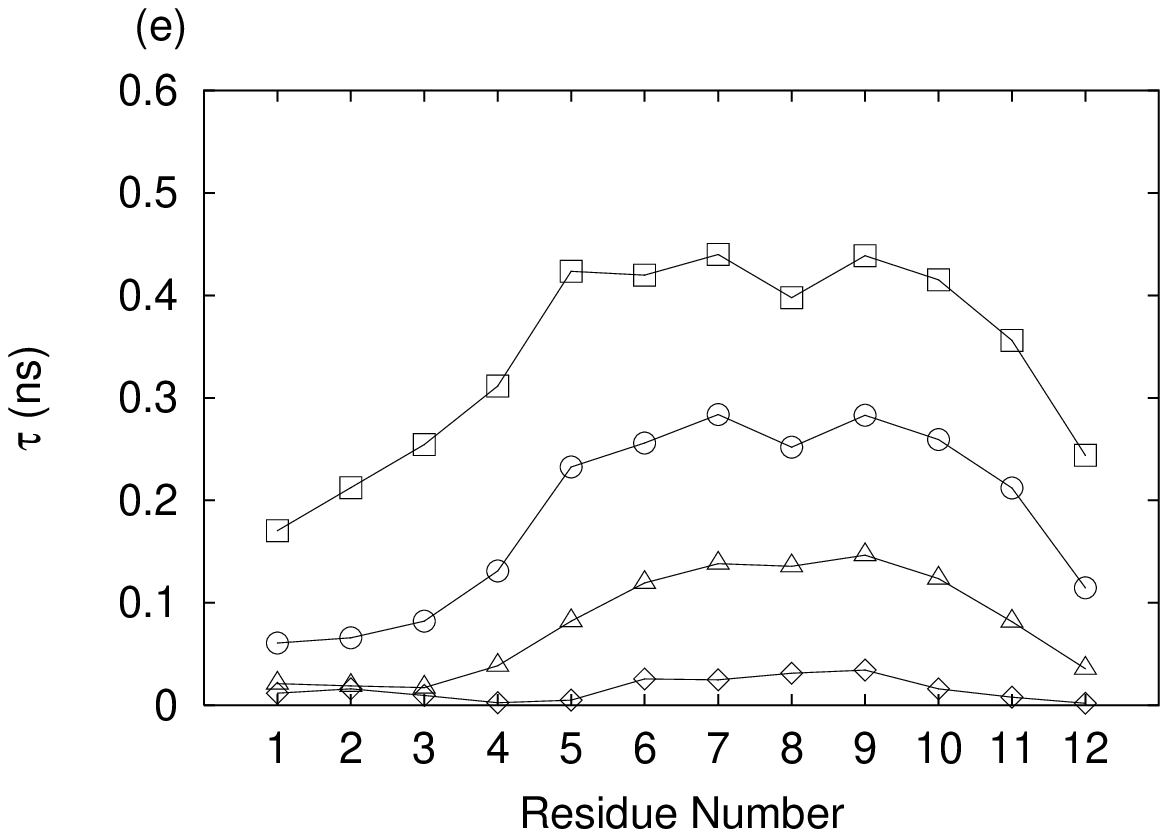}
}

\scalebox{0.7}
{
\includegraphics{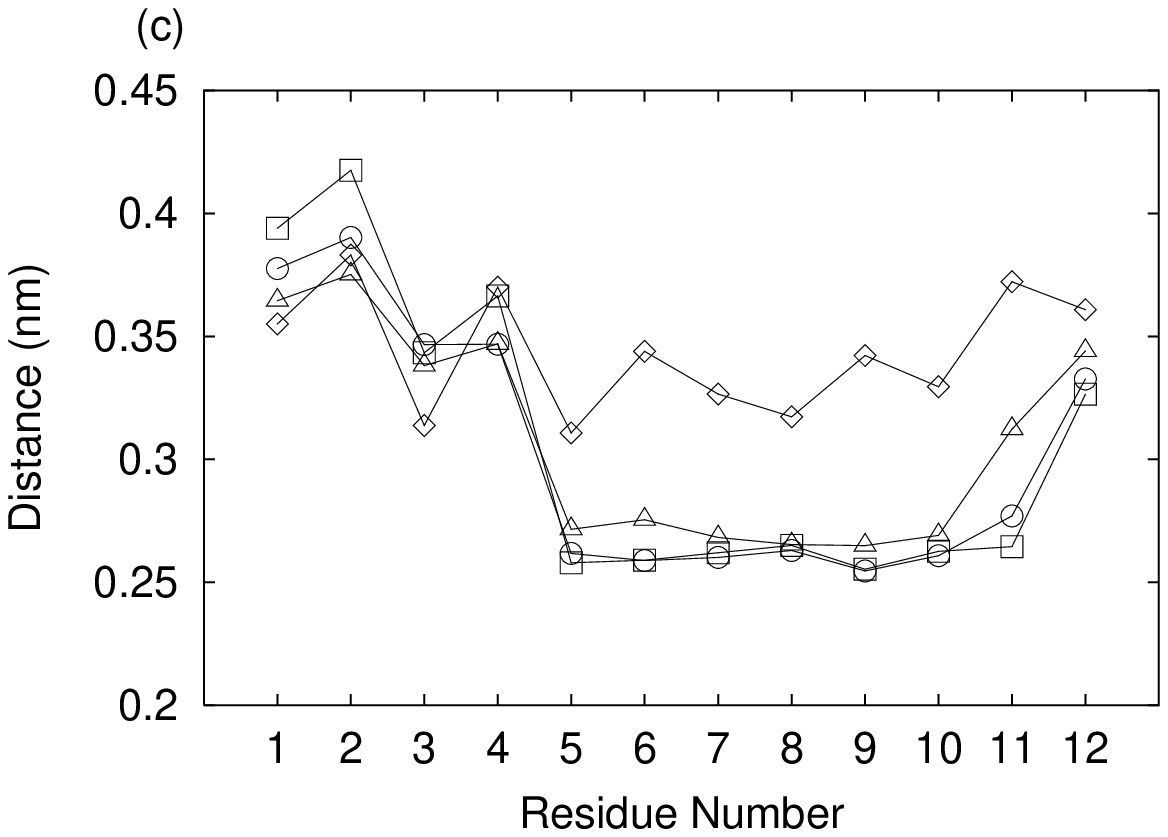}
\hspace{0.2cm}
\includegraphics{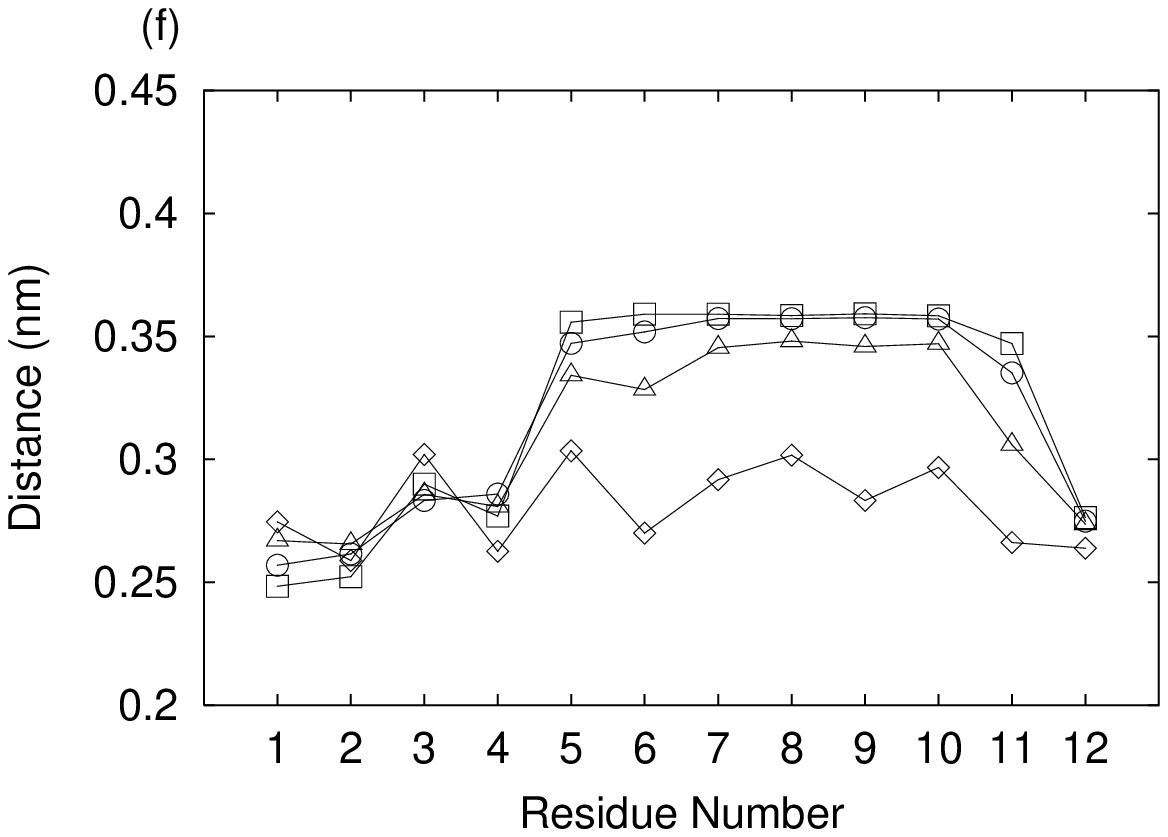}
}
\newpage

\vspace{2cm}
\caption{La Penna et al.}
\label{fig_CPT}
\end{figure}
\newpage

\begin{figure}
\includegraphics{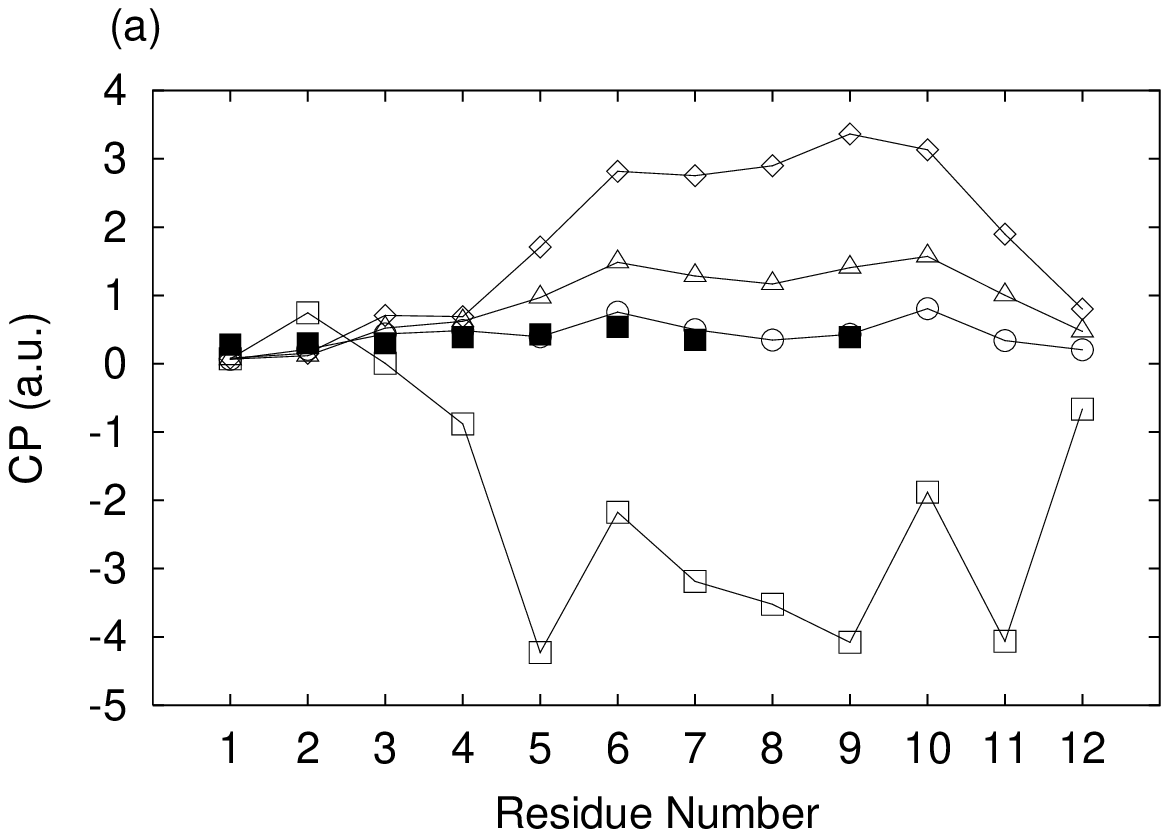}
\hspace{0.2cm}
\includegraphics{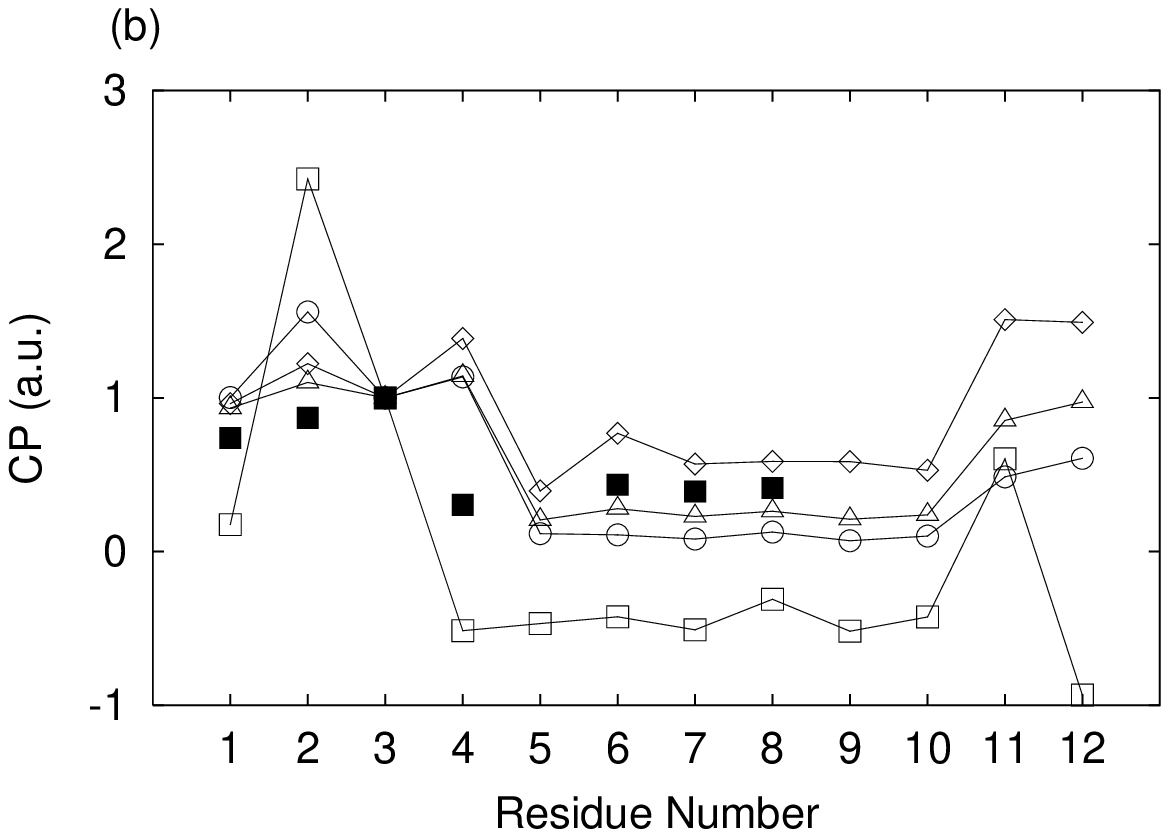}
\vspace{2cm}
\caption{La Penna et al.}
\label{fig_CPexp}
\end{figure}

\end{document}